\documentclass{aa}  

\usepackage{graphicx}
\usepackage{placeins}
\usepackage{txfonts}
\newcommand{\micron}{$\mu$m}              

\usepackage[colorlinks=true,linkcolor=bluea,allcolors=blue]{hyperref}

\begin{document}

   
\title{Silicate emission in a type-2 quasar: JWST/MIRI constraints on torus geometry and radiative feedback}




   \author{C. Ramos Almeida\inst{1,2}, 
A. Asensio Ramos\inst{1,2},
C. Westerdorp Plaza\inst{1,2},
I. García-Bernete\inst{3},
E. Lopez-Rodriguez\inst{4,5},
S. Hönig\inst{6},
A. Audibert\inst{1,2},
S. García-Burillo\inst{7},
M. Pereira-Santaella\inst{8},
F. Donnan\inst{9},
A. Alonso-Herrero\inst{3},
O. Gonz\'alez-Mart\'in\inst{10},
N. Levenson\inst{11},
D. Rigopoulou\inst{12},
C. Tadhunter\inst{13}, 
\and
G. Speranza\inst{8}
          }

   \institute{Instituto de Astrof\' isica de Canarias, Calle V\' ia L\'actea, s/n, E-38205, La Laguna, Tenerife, Spain\\
              \email{cra@iac.es}
              \and Departamento de Astrof\' isica, Universidad de La Laguna, E-38206, La Laguna, Tenerife, Spain
 %
\and Centro de Astrobiolog\' ia (CAB), CSIC-INTA, Camino Bajo del Castillo s/n, E-28692, Villanueva de la Cañada, Madrid, Spain
\and Department of Physics \& Astronomy, University of South Carolina, Columbia, SC 29208, USA
\and Kavli Institute for Particle Astrophysics \& Cosmology (KIPAC), Stanford University, Stanford, CA 94305, USA
\and School of Physics \& Astronomy, University of Southampton, Highfield, Southampton SO171BJ, UK
\and Observatorio Astron\'omico Nacional (OAN-IGN)-Observatorio de Madrid, Alfonso XII, 3, 28014 Madrid, Spain 
\and Instituto de Física Fundamental, CSIC, Calle Serrano 123, 28006 Madrid, Spain
\and Department of Astrophysics, University of California San Diego, 9500 Gilman Drive, San Diego, CA 92093, USA
\and Instituto de Radioastronom\' ia and Astrof\' isica (IRyA-UNAM), 3-72 (Xangari), 8701, Morelia, Mexico
\and Space Telescope Science Institute, Baltimore, MD 21218, USA 
\and Department of Physics, University of Oxford, Oxford OX1 3RH, UK
\and Department of Physics \& Astronomy, University of Sheffield, S3 7RH Sheffield, UK
}

%

   \date{Received Sep 19, 2025; accepted Nov 30, 2025}

 
  \abstract
{Type-2 quasars (QSO2s) are active galactic nuclei (AGN) seen through a significant amount of dust and gas that obscures the central supermassive black hole and the broad line region. Despite this, recent mid-infrared spectra of the central 0.5-1.1 kpc of five QSO2s at z$\sim$0.1, obtained with the MRS module of JWST/MIRI, revealed 9.7, 18, and 23 \micron~silicate features in emission in two of them. This indicates 
that the high angular resolution of JWST/MIRI now allows us to peer into their nuclear region, exposing some of the directly illuminated dusty clouds that produce silicate emission. To test this, we fitted the nuclear mid-infrared spectrum of the QSO2 with the strongest silicate features, J1010, with two different sets of torus models implemented in an updated version of the Bayesian tool {\tt BayesClumpy}. These are the CLUMPY and the CAT3D-WIND models. The CAT3D-WIND model is preferred by the observations based on the marginal likelihood and fit residuals, although the two torus models successfully reproduce the spectrum 
by means of intermediate covering factors ($\rm C_T=0.45\pm^{0.26}_{0.18}$ and $\rm C_T=0.66\pm^{0.16}_{0.17}$ for the CLUMPY and CAT3D-WIND models) and low inclinations ($\rm i=50^\circ\pm^{8^\circ}_{9^\circ}$ and $\rm i=13^\circ\pm^{7^\circ}_{6^\circ}$). Indeed, four of the five QSO2s with JWST/MIRI observations, including J1010, are in the blowout or ``forbidden'' region of the Eddington ratio-column density diagram, indicating that they are actively clearing gas and dust from their nuclear regions, leading to reduced covering factors. This is in contrast with Seyfert 2 galaxies observed with JWST, which are in the ``permitted'' regions of the diagram and show 9.7 \micron~silicate features in absorption. This {\color {black} supports} a scenario where the more luminous the AGN and the higher their Eddington ratio, the lower the torus covering factor, driven by radiation pressure on dusty gas. 
}

   \keywords{galaxies: active -- galaxies: nuclei -- galaxies: quasars -- galaxies:evolution -- ISM: lines and bands}
   
\titlerunning{Silicate emission in a type-2 quasar}
\authorrunning{C. Ramos Almeida et al.}

   \maketitle
%

\section{Introduction}
\label{intro}

Type-2 quasars (QSO2s) are optically selected active galactic nuclei (AGN) with L$_{\rm [OIII]}>$10$^{8.3}L_{\sun}$ that show permitted emission lines with a full width at half maximum (FWHM) of $<$2000 km s$^{-1}$ \citep{Zakamska03,2008AJ....136.2373R}. They are the torus-obscured version of type-1 quasars, as revealed by optical spectropolarimetric data of some QSO2s \citep{2005AJ....129.1212Z}. However, in some cases, part of the obscuration might come from galactic scales \citep{Polletta08,Fawcett23}, and indeed, they might be a dust-embedded phase during which AGN-driven outflows start clearing up gas and dust to eventually become unobscured quasars \citep{1988ApJ...325...74S,2009ApJ...696..891H,Lansbury20,Hauschild25,Molyneux25}. 

High angular resolution infrared observations are key to characterize nuclear dust, since, as mentioned above, obscuration in AGN might not only be associated with the torus \citep{Ramos17review}, but also with dust in the host galaxy. 
Spectra from the Spitzer Space Telescope, for example, correspond to large apertures ($\sim$3.6-10.2\arcsec) that, even in the case nearby targets, include a significant fraction of dust within the host galaxy. Although spectral decomposition techniques proved to work nicely in isolating the nuclear AGN component to then fit it with torus models (e.g., \citealt{GonzalezMartin19,GonzalezMartin23}), now, thanks to the James Webb Space Telescope (JWST) it is possible to obtain mid-infrared spectra of AGN in the range 5-28 \micron~with a resolution of $\sim$0.3-0.8\arcsec. This allows us to probe the central tens of parsecs of nearby AGN to study torus properties \citep{GarciaBernete24,SODA2024,Haidar24,Gonzalez25}.

The strengths of the silicate features detected in the mid-infrared depend on the geometry of the dusty clouds, while their profile shapes depend on the dust properties, such as porosity, grain size, and composition \citep{Li08,Thompson09,Henning10,Honig10,Smith10,GonzalezMartin23,Reyes24,Reyes25}. In particular, the 9.7 \micron~silicate feature strength (S$_{\rm 9.7}$), widely measured in AGN using data from the InfraRed Spectrograph (IRS) of Spitzer, is known to correlate with AGN type and gas column density (N$_{\rm H}$; \citealt{Shi06,Hatziminaoglou15}), with obscured, type-2 AGN generally showing it in  absorption and unobscured, type-1 AGN either in emission or flat. However, examples of type-2 AGN with silicate features in emission and of type-1 AGN with silicates in shallow absorption have been reported in the literature using data from Spitzer/IRS (e.g., \citealt{Hatziminaoglou15}), and from subarcsecond resolution mid-infrared spectroscopic data obtained with ground-based telescopes \citep{Mason09,Honig10,Alonso16,MartinezParedes17}. Indeed, these deviations from the general trend constituted one of the motivations for the development of clumpy torus models (see \citealt{Ramos17review} for review), whose silicate features (either in absorption or emission) are always relatively weak because both illuminated and shadowed cloud sides contribute to them. While most views of the torus in type-2 AGN intercept absorbing shadowed cloud faces, silicate emission from some bright cloud faces fills in the feature, making it shallower, flat, or in emission \citep{Nenkova08a,Nenkova08b}.

\citet{Zakamska16} 
reported Spitzer/IRS S$_{\rm 9.7}$ measurements for a sample of 46 type-2 AGN, including QSO2s, at a median redshift of z=0.17. {\color {black} Only one of the targets shows the silicate feature in relatively strong emission (S$_{9.7}$=0.27), although it is classified as a type-1 AGN in \citet{Veron06} and \citet{Lyu18}. The other 45 type-2 AGN in \citet{Zakamska16} have S$_{9.7}$=[-2.5,0], with a median value of $-0.30$.} 
\citet{Sturm06} reported tentative evidence of 9.7 \micron~silicate in weak emission or absent in the spectra of five obscured quasars with redshifts between 0.2 and 1.4, but based on Spitzer/IRS spectra covering a narrow spectral range and without subtracting polycyclic aromatic hydrocarbon (PAH) emission. 

The first mid-infrared JWST spectroscopic observations of a sample of nearby QSO2s (with z$\sim$0.1 and log L$_{\rm bol}\sim$45.5 erg~s$^{-1}$) revealed striking differences in their continuum shapes and silicate feature strengths (see Fig. 1 in \citealt{Ramos25}). Two of the QSO2s, J1010 and J1100, have S$_{9.7}$=0.49$\pm$0.01 and 0.11$\pm$0.01 (emission) that did not change after PAH removal, while the other three show absorption features with S$_{9.7}$ ranging from $-0.2$ to $-1.0$ ($-0.05$ to $-0.95$) before (after) PAH subtraction. The amorphous silicate feature at 18 \micron~and the crystalline silicate band at 23 \micron~were detected in emission in three of the QSO2s (J1010, J1100, and J1430), consistent with relatively low obscuration \citep{Spoon22,GarciaBernete24}. Indeed, the gas column densities measured from molecular gas observations \citep{Ramos22} range from log N$_{\rm H}^{\rm CO}\sim$22 to 23 cm$^{-2}$, which are modest values considering that the targets are obscured quasars. Only in two of the QSO2s, J1356 and J1509, which are two of the most obscured ones (log N$_{\rm H}^{\rm CO}\sim$23 cm$^{-3}$), the 18 \micron~silicate feature is in weak emission, there is no 23 \micron~silicate feature, and there are absorption bands of CO and H$_2$O ices and aliphatic grains \citep{Ramos25}.


This diversity in the silicate strengths of the QSO2s is at odds with the recent results found for Seyfert galaxies, also based on JWST mid-infrared spectra \citep{GarciaBernete24,Gonzalez25}. The nuclear spectra of six Seyfert 2 galaxies from the Galaxy Activity, Torus and Outflow Survey (GATOS), which have X-ray column densities of log N$_{\rm H}$=22.2-24.3 cm$^{-2}$ and bolometric luminosities of log L$_{\rm bol}\sim$43.5-44.5 erg~s$^{-1}$, show the 9.7 and 18 \micron~silicate features in absorption, {\color {black} absorption bands of ices and aliphatic grains, and no 23 \micron~silicates} \citep{GarciaBernete24}. 
More recently, \citet{Gonzalez25} compiled JWST/MIRI spectra of a sample of 21 nearby (z<0.05) AGN (7 type-1 and 14 type-2), including the six Seyfert 2 galaxies studied by \citet{GarciaBernete24}. {\color {black} These AGN have log L$_{\rm bol}\sim$40-45 erg~s$^{-1}$ (estimated by multiplying by 20 the intrinsic L$_{\rm 2-10~keV}$ values there reported), except the quasar Mrk~231, which has log L$_{\rm bol}>$45 erg~s$^{-1}$.} They found all the type-2 AGN in their sample showing 9.7 \micron~silicates in absorption, some of them deep, with their spectra showing water ices and aliphatic grains in absorption, indicative of high obscuration. These spectra correspond to smaller spatial scales than those of the QSO2s because of their redshifts (z<0.05 for the Seyferts versus z$\sim$0.1 for the QSO2s). 

Here we attempt to reproduce the JWST mid-infrared spectrum of the QSO2 with the strongest silicate emission features in \citet{Ramos25}, J1010, with torus models. The combination of spectral coverage and angular resolution of JWST/MIRI is excellent for constraining the model parameters, according to previous work  \citep{Ramos14,GonzalezMartin19,Gonzalez25}. To do so we developed \texttt{BayesClumpy2}, an updated version of the Bayesian tool described in \cite{Asensio09,Asensio13}, and tested whether the two sets of torus models implemented on it can reproduce the nuclear mid-infrared spectrum of this QSO2 with strong silicate emission. The models are the clumpy torus model of \citet{Nenkova08a}, also known as CLUMPY, and the disc+wind model of \citet{Hoenig17}, CAT3D-WIND. We then compare JWST/MIRI measurements of the 9.7 \micron~silicate feature available in the literature for Seyfert 2 galaxies and QSO2s \citep{GarciaBernete24,Ramos25} with different AGN and galaxy properties, to investigate the origin of the difference between their obscuration properties. In the following, we assume a cosmology with H$_0$=70 km~s$^{-1}$ Mpc$^{-1}$, $\Omega_m$=0.3, and $\Omega_{\Lambda}$=0.7.

\section{Target and mid-infrared continuum spectrum}
\label{observations}

\begin{figure}
  \includegraphics[width=1.05\columnwidth]{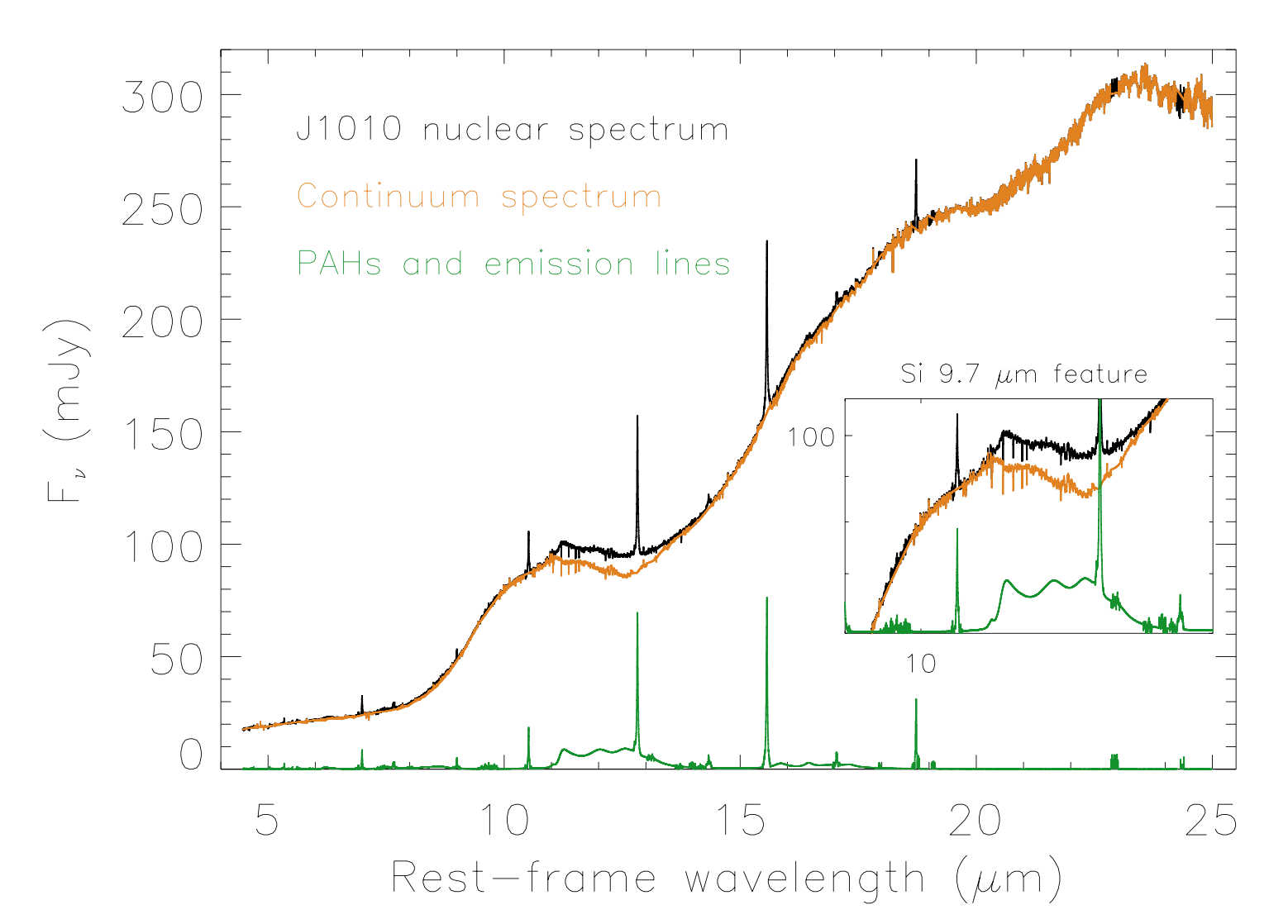}
   \caption{Nuclear mid-infrared spectrum of J1010 from \citet{Ramos25} (in black), and continuum spectrum (in orange) obtained from subtracting the PAH bands and the most prominent emission lines (in green) from the nuclear spectrum. The inset shows the 9.7 \micron~silicate feature in logarithmic scale.}
              \label{fig1}%
    \end{figure}

The QSO2 studied here, {\color {black} J1010 (SDSS J101043.36+061201.4),} is part of the Quasar Feedback (\href{http://research.iac.es/galeria/cra/qsofeed/}{QSOFEED}) sample \citep{Ramos22,Pierce23,Bessiere24}, and its nuclear mid-infrared spectrum was first presented in \citet{Ramos25}. It was observed in May 22, 2024 with the integral field unit of JWST/MIRI, the MRS, as part of Cycle 2 General Observer (GO) Program 3655 (PI: C. Ramos Almeida; MAST \href{https://archive.stsci.edu/doi/resolve/resolve.html?doi=10.17909/8w9h-re72}{{doi:10.17909/8w9h-re72}}). 
We refer the reader to the Program Information webpage of Program \href{https://www.stsci.edu/jwst/science-execution/program-information?id=3655}{{GO 3655}} and to \citet{Ramos25} for further details on the MRS observations and data reduction. The MIRI/MRS nuclear spectrum covers the rest-frame spectral range $\sim$5-25 $\rm{\mu m}$, with a spectral resolution of R$\sim$3700-1300 \citep{2021A&A...656A..57L,Argyriou23}.
This nuclear spectrum was extracted from all the MRS sub-channels assuming that it is a point source, using apertures ranging from $\sim$0.3\arcsec~to 0.8\arcsec~with increasing wavelength, which corresponds to physical scales of 0.5-1.1 kpc at the redshift of the target (z=0.0977). The spectra of the 12 bands were then stitched together using small scaling factors (between 0.97 and 1.05) to make them coincide in the overlapping regions (see \citealt{Ramos25} for further details).  
We then used the fit performed with the tool described in \citet{Donnan24}, which models the PAH features on top of a continuum, to subtract them and the most prominent emission lines detected in the nuclear spectrum, obtaining the continuum spectrum shown in Figure \ref{fig1}.

We selected J1010 for this study because, as can be seen from Figure \ref{fig1}, it is continuum-dominated in the mid-infrared, with a relatively small contribution from PAHs, and despite its type-2 classification in the optical (see below), it shows the 9.7, 18, and
23 \micron~silicate features in emission \citep{Ramos25}. In fact, the strength of its 9.7 \micron~silicate feature, of S$_{\rm 9.7}$=0.49$\pm$0.01, is among the 10 highest values measured for the type-1 AGN in \citet{Mariela20}, which range between 0.3 and 0.5. This indicates that we are seeing emission from hot dust, heated to temperatures of several hundred kelvin to $\sim$1000 K \citep{Li08}, coming from the central 0.5-1.1 kpc of J1010, with little obscuration from colder dust. {\color {black} It is therefore an interesting target to test whether state-of-the-art torus models can reproduce its nuclear mid-infrared spectrum without including the broken power-law that it is usually fitted in the case of type-1 AGN to account for direct emission from the accretion disc} \citep{Ramos09,Ramos11,GarciaBernete19}. See Table \ref{tab1} for information on the AGN and galaxy properties of J1010.

All the QSOFEED QSO2s, including J1010, were selected from the \citet{2008AJ....136.2373R} compilation of narrow-line AGN, from where type-1 AGN were removed. 
However, as discussed in \citet{Ramos25}, the H$\alpha$+[NII] profile detected in the {\color {black} continuum-subtracted} optical SDSS spectrum of J1010 appears slightly broader than that of H$\beta$ (see Figs. A1, {\color {black} A2, and A3} in \citealt{Ramos25}), which could be consistent with a type 1.9 AGN classification instead of type-2 \citep{Osterbrock81}. To test this possibility, \citet{Ramos25} 
performed a multi-component Gaussian fit of the H$\beta$, [OIII], and H$\alpha$+[NII] lines {\color {black} (see Fig. A3 there). They} found that, while the H$\beta$ and [OIII] profiles could be successfully reproduced with four Gaussian components (some of them corresponding to outflowing gas; \citealt{Bessiere24,Speranza24}), the inclusion of a fifth Gaussian component improved the reduced $\chi^2$ of the H$\alpha$+[NII] fit by 86\%. This fifth component has a FWHM=3400$\pm$100 km~s$^{-1}$ and it is blueshifted by 1000$\pm$70 km~s$^{-1}$ relative to systemic. {\color {black} Based on this} large blueshift, \citet{Ramos25} argued that it is more likely to be associated with an additional, weak outflow component than with the broad line region (BLR). Therefore, the more recent analysis of the SDSS spectrum agrees with the type-2 classification done by \citet{2008AJ....136.2373R} and we consider it as such hereafter. {\color {black} Nevertheless, it is noteworthy that the nuclear column density derived from CO data ($\rm log~N_{H}^{CO}=22.3~cm^{-2}$; see Table \ref{tab1}) places J1010 in the boundary between type 1.8-1.9 and type-2 AGN, according to \citet{Burtscher16}. Assuming $\rm N_H/A_V=[1.8,2.7]\times10^{21}~cm^{-2}$ \citep{Predehl95,Nowak12}, this corresponds to $\rm A_V\sim[7,11]$ mag. To estimate $\rm A_V^{broad}$ from the Balmer decrement we unsuccessfully attempted to force a fifth Gaussian component in the H$\beta$ fit reported by \citet{Ramos25} by fixing the kinematics to those of H$\alpha$ and allowing the line amplitudes to vary. We then calculated an upper limit at 2$\sigma$ for the flux of that missing H$\beta$ broad component. Using that upper limit, the H$\alpha$ flux of the broad component reported in Table A.1 in \citet{Ramos25}, and the \citet{Cardelli89} extinction law, we get A$_V$>5 mag.}

\begin{table}
\centering
  \caption{Properties of the QSO2 J1010 (SDSS J101043.36+061201.4).} 
\begin{tabular}{lc}
\hline
\hline
SDSS z                                   & 0.0977 \\
D$_{\rm L}$ (Mpc)                              & 449 \\
Scale (kpc/arcsec)                                   & 1.807 \\
A$_{\rm V}$ (mag)                               & 1.1 \\
log L$_{\text{bol}}$  (erg s$^{-1}$)                   & 45.55 \\
log L$_{1.4\text{GHz}}$  (W Hz$^{-1}$)                & 24.37 \\
log M$_{\rm BH}$  (M$_{\odot}$)                        & 8.4$\pm$0.8 \\
log\(\frac{L_{\rm bol}}{L_{\rm Edd}}\)   &-0.8$\pm$0.8\\ 
log M$_*$     (M$_{\odot}$)                            & 11.0$\pm$0.2\\
SFR   (M$_{\odot}$yr$^{-1}$)                                   & 32, 34, 7\\
log N$_{\rm H}^{\rm CO}$ (cm$^{\rm -2}$) & 22.3 \\ 
\hline
\end{tabular}
\tablefoot{A$_{\rm V}$ derived from H$_{\alpha}$/H$_{\beta}$ reported by \citet{Kong18} and the extinction law from \citet{Cardelli89}. Bolometric luminosity calculated by applying the correction factor of 474 from \citet{Lamastra09} to the extinction-corrected [O~III]5007 \AA~ luminosity from \citet{Kong18}, and 1.4 GHz luminosity and stellar mass from \citet{Ramos22}. BH mass and Eddington ratio are from \citet{Kong18}. The SFRs were derived from the total IR luminosity from \citet{Ramos22}, from spectral synthesis modelling of the SDSS spectra performed by \citet{Bessiere24}, and from the nuclear 11.3 \micron~PAH detected in the JWST/MIRI spectrum from \citet{Ramos25}. The gas column density comes from \citet{Ramos25} and it was measured
from the ALMA CO(2-1) observations studied in \citet{Ramos22}.} 
\label{tab1} 
\end{table}

\section{Bayesian inference with torus models}
\label{bayesclumpy}

In \citet{Asensio09} we presented {\tt BayesClumpy}, a tool that permits the user to efficiently carry out Bayesian analysis of observed spectral energy distributions (SEDs) including photometric points and/or spectra using the clumpy torus models of \citet{Nenkova08a}. This tool is based on a Markov Chain Monte Carlo code whose output is the posterior distribution function (PDF) for all the parameters of the clumpy models once the observations are taken into account. As a consequence, the code yields statistically significant estimations of the parameters and, more important, statistically relevant confidence intervals. This has been proved to be a suitable approach for overcoming the degeneracies associated with these models, based on eight parameters if we include the vertical shift necessary to scale the models to the observations and the foreground extinction (see Table \ref{tab:nenkova}). Several works have used {\tt BayesClumpy} to fit observational data with these torus models, making them more accessible to the community (e.g., \citealt{Ramos09,Ramos11,Ramos14,Alonso11,Ruschel14,Ichikawa15,Mateos16,Gonzalez17,Audibert17,MartinezParedes17,GarciaBernete15,GarciaBernete19}).

\begin{table*}
\centering
\small
\caption{Parameters of the clumpy torus models of \citet{Nenkova08a}.}
\begin{tabular}{lccccc}
\hline
\hline
Parameter 									& Symbol 		& Uniform prior 	&  \multicolumn{3}{c}{Posterior percentiles} \\ 
                                            &              &                    &  16   &   50   &  84 \\ 
\hline
Radial extent  				             &	Y		 &	[5, 100]		         &  23 & 26  & 30   \\ 
Width of clouds angular distribution     &	$\sigma$ &	[15$^\circ$, 70$^\circ$] &  28$^\circ$ & 35$^\circ$  & 43$^\circ$   \\ 
Number of clouds along an equatorial ray &	N$_0$	 &	[1, 15]		             &   6 &  8  & 12   \\ 
Index of the radial density profile	(r$^{\rm -q}$)	 &	q	     &	[0, 3]		             & 0.07& 0.26& 0.57 \\ 
Optical depth per single cloud		     &	$\tau_V$ &	[10, 300]		         & 233 & 269 & 292  \\ 
Inclination			                     &	i	     &	[0$^\circ$, 90$^\circ$]	 &  41$^\circ$ & 50$^\circ$  & 58$^\circ$   \\ 
Foreground extinction (mag)              &   A$_V$   &   [0,2]                   & 0.4 & 1.0 & 1.6  \\ 
\hline
Covering factor                          &  C$_T$    &              \dots            & 0.27 & 0.45 & 0.70 \\
Type-2 probability                       & 1-P$_{\rm esc}$    &    \dots              & 0.24 & 0.89 & 1.00 \\
Torus extinction (mag)                   & A$_V^{\rm LOS}$    &   \dots                & 71 & 633 & 2187 \\
\hline
\end{tabular}					 
\tablefoot{i$=$0 is face-on and i$=$90 is edge-on. The foreground extinction, A$_V$, is an additional parameter that we use, together with the extinction curve of \citet{Chiar06}, to take into account the foreground extinction from the host galaxy. These models use a standard Galactic mix of 53\% silicates and 47\% graphites, with maximum sizes of 0.25 \micron. {\color {black} $\rm C_T=1-\int e^{-N_{LOS}(i)}d cos(i)$, $\rm P_{esc}=e^{-N_{LOS}(i)}$, and $\rm A_{V}^{LOS} = 1.086~\tau_{V}~N_{LOS}(i)$, with $\rm N_{LOS}(i) = N_0~e^{-(90-i)^2/\sigma^2}$. C$_T$ is an integrated quantity (over the inclination angle, i), whilst P$_{\rm esc}$ and A$_V^{\rm LOS}$ are calculated at a given i.}}
\label{tab:nenkova}
\end{table*}

\begin{table*}
\centering
\small
\caption{Parameters of the disc+wind models of \citet{Hoenig17}.}
\begin{tabular}{lccccc}
\hline
\hline
Parameter 									& Symbol 		& Uniform prior 	&  \multicolumn{3}{c}{Posterior percentiles} \\
                                            &               &                   &  16   &   50   &  84  \\
\hline
Radial distribution power law index disc (r$^{\rm a}$)	& a	  & [-3.0,-0.5]		        & -1.8 & -1.7 & -1.5 \\
Number of clouds along an equatorial ray    & N$_0$           & [5,10]		            &    6 & 8    & 9 \\
Scale height of the disc		            & h	              & [0.1,0.5]		        & 0.13 & 0.17 & 0.22\\
Radial distribution power law index wind (r$^{\rm a}$)& a$_{\rm w}$	& [-2.5,-0.5]		& -0.8 & -0.7 & -0.6 \\
Half-opening angle of the wind		        & $\theta_{\rm w}$& [30$^\circ$,45$^\circ$] &   37$^\circ$ & 40$^\circ$   & 43$^\circ$ \\
Angular width of the hollow wind cone		& $\sigma$	      & [7$^\circ$,15$^\circ$]	&   10$^\circ$ & 12$^\circ$   & 14$^\circ$ \\
Wind-to-disc ratio of dust clouds		    & f$_{\rm wd}$	  & [0.15,2.25]		        &  1.5 & 1.8  & 2.1 \\
Outer radius                                & r$_\mathrm{out}$& [450,500]               &  471 & 484  & 494\\
Inclination 				                & i	              & [0$^\circ$,90$^\circ$]	&    7$^\circ$ & 13$^\circ$   & 20$^\circ$ \\
Foreground extinction (mag)                 & A$_V$           & [0,2]                   &  0.2 & 0.5  & 1.0\\ 
\hline
Covering factor                          &  C$_T$             &   \dots           & 0.49 & 0.66 & 0.82 \\
Type-2 probability                       & 1-P$_{\rm esc}$    &   \dots           & 0.00 & 0.12 & 0.95 \\
Torus extinction (mag)                   & A$_V^{\rm LOS}$    &      \dots             & 0.01 & 7 & 160 \\
\hline
\end{tabular}					 
\tablefoot{i$=$0 is face-on and i$=$90 is edge-on. {\color {black} The optical depth per cloud is $\tau_V$=50, and A$_V$ is defined as treated as} in Table \ref{tab:nenkova}. These models use a mixture of the standard ISM dust composition of 53\% silicates and 47\% graphites with maximum sizes of 1 \micron~in the disc, plus a graphite-dominated population of small grains (sizes $\leq$0.25 \micron), and large graphites only in the wind (0.75-1 \micron). {\color {black} $\rm C_T$, $\rm P_{esc}$, and $\rm A_{V}^{LOS}$ as in Table \ref{tab:nenkova}, but with $\rm N_{LOS}(i) = N_0~f_{wd}~e^{-((90-i)-\theta_w)^2/2 \sigma^2} + N_0~e^{-(tan(90-i))^2/2 h^2}$.}}
\label{tab:honig}
\end{table*}

Based on mid-infrared interferometry data of nearby AGN, which revealed elongated polar emission on scales of 1-10 pc in addition to a more compact equatorial disc component \citep{Hoenig12,Hoenig13,Tristram14}, \citet{Hoenig17} developed a clumpy torus model including a disc and a wind. The latter consists of a hollow cone that allows the observer to see type-1 AGN views (see \citealt{Hoenig19} for a detailed description of the two components at different wavelengths). These models have 11 free parameters including vertical shift and foreground extinction (see Table \ref{tab:honig}) and they have been successfully fitted to the SEDs of nearby AGN \citep{GonzalezMartin19,GonzalezMartin23,Gonzalez25,Esparza21,GarciaBernete22torus}. The results indicate that these models are statistically preferred by the observations in the case of luminous type-1 AGN, while clumpy torus models are the best choice to reproduce the SEDs of lower-luminosity type-2 AGN \citep{GonzalezMartin19,GarciaBernete22torus}.

With the aim of incorporating the disc+wind models of \citet{Hoenig17} into {\tt BayesClumpy}, we have developed a new version of the tool, {\tt BayesClumpy2}\footnote{\href{https://github.com/aasensio/bayesclumpy2}{\texttt{https://github.com/aasensio/bayesclumpy2}}.}, that is available to the community. 
To maintain backward compatibility, the \cite{Nenkova08a} models are 
interpolated using a multilinear interpolator, as in the original version of {\tt BayesClumpy}. However, since the number of parameters of the \citet{Hoenig17} models 
is larger, for them we used a neural interpolator. To this end, we use the same methodology outlined in \citet{2023A&A...675A.191W}. In the case of the disc+wind models the dimensionality of the data (torus models of 105 points in wavelength) was reduced by applying a Convolutional Auto Encoder (CAE) with a bottleneck of 32 neurons. On the resulting 32 embeddings we then trained a custom neural network (NN). This custom NN consists of 32 layers, each with 128 neurons, fully connected and propagating the residuals skipping 2 layers. 

\texttt{Bayesclumpy2} is now a Python code with some Fortran 90 code to accelerate multilinear interpolations. {\color {black} It} is controlled via human-readable configuration files that are used to: 1) select the set of models to be used in the inference; 2) change the specific MCMC sampler; 3) include (for type-1 AGN) or not (for type-2 AGN) {\color {black} direct emission from the accretion disc by means of a broken power-law}; and 4) select the  extinction law to be applied in combination with the foreground extinction (A$_V$). The observations are entered via a text file containing photometric and/or spectroscopic data. Finally, since \texttt{Bayesclumpy2} carries out Bayesian inference, it requires the specification of the a priori information about the model parameters in the fit. The prior distributions can be uniform, normal, or Dirac, depending on the information available.

\begin{figure*}
   \centering
\includegraphics[width=1.9\columnwidth]{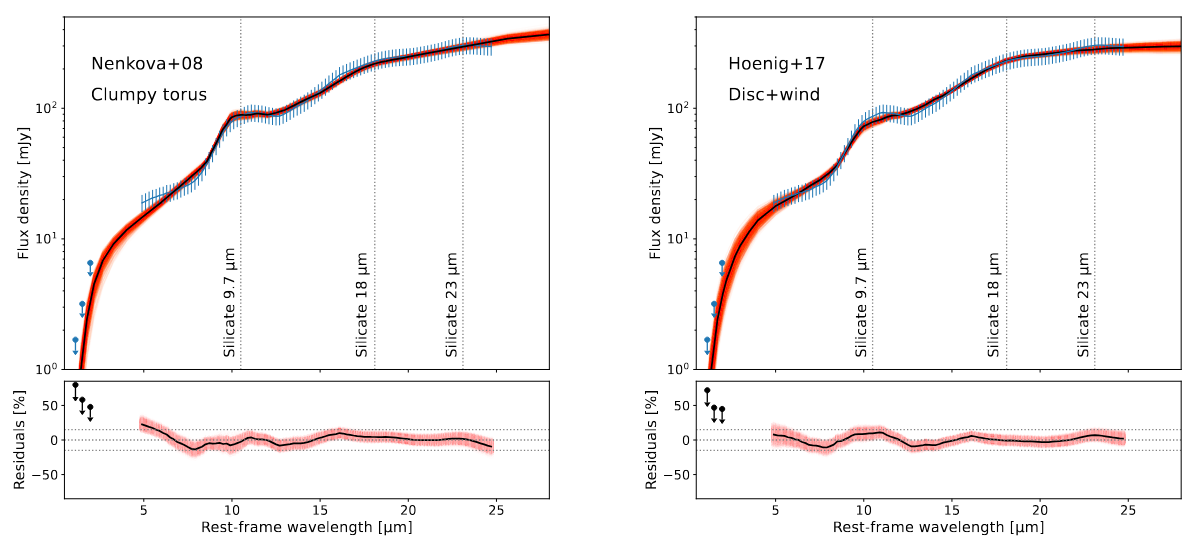}
   \caption{Smoothed and interpolated nuclear spectrum of J1010, with the emission lines and the PAH features removed (solid blue line), and best fits with torus models (shaded orange area, which corresponds to model SEDs between the 16 and 84 percentiles of the posterior and solid black line corresponding to the 50 percentil of the posterior). The left panels correspond to the fit with the clumpy torus models of \citet{Nenkova08a}, and the right panel to the disc+wind models of \citet{Hoenig17}. Vertical dotted lines are the peak wavelengths measured for the silicate features in the nuclear spectrum (see Table \ref{tab:silicates}). The bottom panels show the residuals as a percentage of the continuum flux, with the horizontal dotted lines indicating 0 and 15\% error for guidance. {\color {black} The blue and black upper limits are the J, H, and Ks fluxes from the 2MASS Point Source Catalog (1.69, 3.18, and 6.55 mJy, respectively) and corresponding residuals. These upper limits are shown for comparison only and were not included in the fit.}}
              \label{fig2}%
    \end{figure*}

Here we used a smoothed and interpolated version of the nuclear continuum spectrum of J1010 {\color {black} shown in Fig. \ref{fig1}} (i.e., after PAH and emission line removal). We smoothed the spectrum using a Lee filter algorithm with a box size of 100, and then interpolated it to a grid of 100 wavelengths that successfully captured the spectral shape, including the silicate features. The smoothed and interpolated version of the nuclear mid-infrared spectrum is shown in Fig. \ref{fig2}. According to the fit with the infrared tool presented by \citet{Donnan24}, the nuclear mid-infrared spectrum of the QSO2 is continuum-dominated, with a small contribution from PAHs (see Fig. \ref{fig1}). This is consistent with the observed K-band excess reported for this source by \citet{Shangguan19} and \citet{Jarvis20}, indicative of an important contribution from AGN-heated dust \citep{Mor09,Ramos09b,Ramos11b,GarciaBernete17}. Therefore, it is reasonable to assume that the bulk of this nuclear spectrum (probing the inner 0.5-1.1 kpc of the QSO2), including the silicate features, comes from the torus, with minimal contribution from the host galaxy. This has been independently confirmed by using the recently developed tool {\it MRSPSFisol}, that decomposes MIRI/MRS cubes into nuclear and extended contributions \citep{Gonzalez25}. 

We run the fits with {\tt BayesClumpy2} using one of the nested sampling \citep{2004AIPC..735..395S} samplers available, no direct emission from the accretion disc because it is a type-2 AGN, and the extinction law of \citet{Chiar06}. In the case of J1010, the only a-priori information we have is the foreground extinction, A$_V$=1 mag, calculated from the Balmer decrement by \citet{Kong18}. Therefore we use a uniform prior between 0 and 2 mag to account for some uncertainty (see Tables \ref{tab:nenkova} and \ref{tab:honig}). The rest of priors are uniform using the full parameter ranges available within the models, which are reported in the corresponding tables. We assumed 15\% of error for the nuclear spectrum, which is another input to \texttt{Bayesclumpy2}, to account for the flux calibration uncertainty ($\sim$10\%; \citealt{Ramos25}) and the error associated with the extraction of the nuclear spectrum. 

\section{Results}
\label{results}

Here we show the results of the fit of the nuclear mid-infrared spectrum of J1010 using the two sets of torus models. These results include the posterior or probability distributions, the model images, and the model spectra. 

\subsection{Posterior distributions}

The solutions to the Bayesian inference problem are the posterior or probability distributions resulting for each parameter, which we show in Figures \ref{posteriors_nenkova} and \ref{posteriors_hoenig}. The 2D posteriors are useful to inspect possible degeneracies between pairs of parameters, and the 1D distributions show how much the observations have managed to constrain the model parameters. The 16, 50, and 84 percentiles of these 1D posteriors are shown as vertical dashed lines in the plots, and the corresponding values are shown in Tables \ref{tab:nenkova} and \ref{tab:honig}. As can be seen from Figures \ref{posteriors_nenkova} and \ref{posteriors_hoenig}, the MIRI/MRS spectrum successfully constrains the model parameters, which differ significantly from the uniform priors. In the case of the clumpy torus model, we have a torus median model (50 percentile) with small extent (Y=26), intermediate width ($\sigma$=35$^{\circ}$), number of clumps (N$_0$=8), and inclination (i=50$^{\circ}$). All these posteriors are Gaussian-shaped and the median value (50 percentile) is representative of the posterior peak. The index of the radial cloud distribution, $q$, and the optical depth of the clouds, $\tau_V$, on the other hand, are concentrated towards the lowest and highest values or the parameter interval, respectively, indicating a flat cloud distribution and high optical depth.  

In the case of the disc+wind model, which is described by a larger number of parameters, most of the posteriors are also Gaussian-like, as e.g., the radial distribution power law index of the disc and the number of clouds, which have intermediate values of a=$-1.7$ and N$_0$=8 (values at the 50 percentile). The outer radius of the torus is large, with a median of R=484, and the inclination close to face-on, with i=13$^{\circ}$. The wind half opening angle and angular width show high values of $\sigma_w$=40$^{\circ}$ and $\sigma$=12$^{\circ}$, respectively. Finally, the wind-to-disc ratio of clouds has an intermediate value of f$_{\rm wd}$=1.8, and the scale height of the disc, h, and the radial distribution power law index of the wind, a$_w$, are concentrated towards the lowest and highest values within the parameter interval, respectively. This means a thin disc and a wind with a flat cloud distribution. We refer the reader to Fig. 1 in \citet{GarciaBernete22torus} for a schematic view of the parameters of the two sets of models. 

Since the error that we are assuming for the nuclear spectrum is relatively large (15\%), we evaluated the fits when errors of 10 and 5\% are used instead. We found that using 10 and 15\% error produces practically the same posterior distributions, slightly narrower in the case {\color {black} of the former for some of the} parameters. When we use 5\% error, the posteriors that are Gaussian {\color {black} become} significantly narrower, albeit centered at the same values as those shown in Tables \ref{tab:nenkova} and \ref{tab:honig}, {\color {black} and} those which are not Gaussian (e.g., q and $\tau_V$ in the clumpy model) have median values peaking at the edge of the prior intervals. From this analysis we conclude that a 10-15\% error is adequate to fit torus models to MIRI/MRS spectra using Bayesian statistics. Smaller errors can lead to solutions that are too restrictive in the range of parameters consistent with the data, not permitting the Bayesian inference to correctly sample the posterior distribution \citep{Asensio09,Asensio13}.  

  \begin{figure*}
   \centering
  \includegraphics[width=2.0\columnwidth]{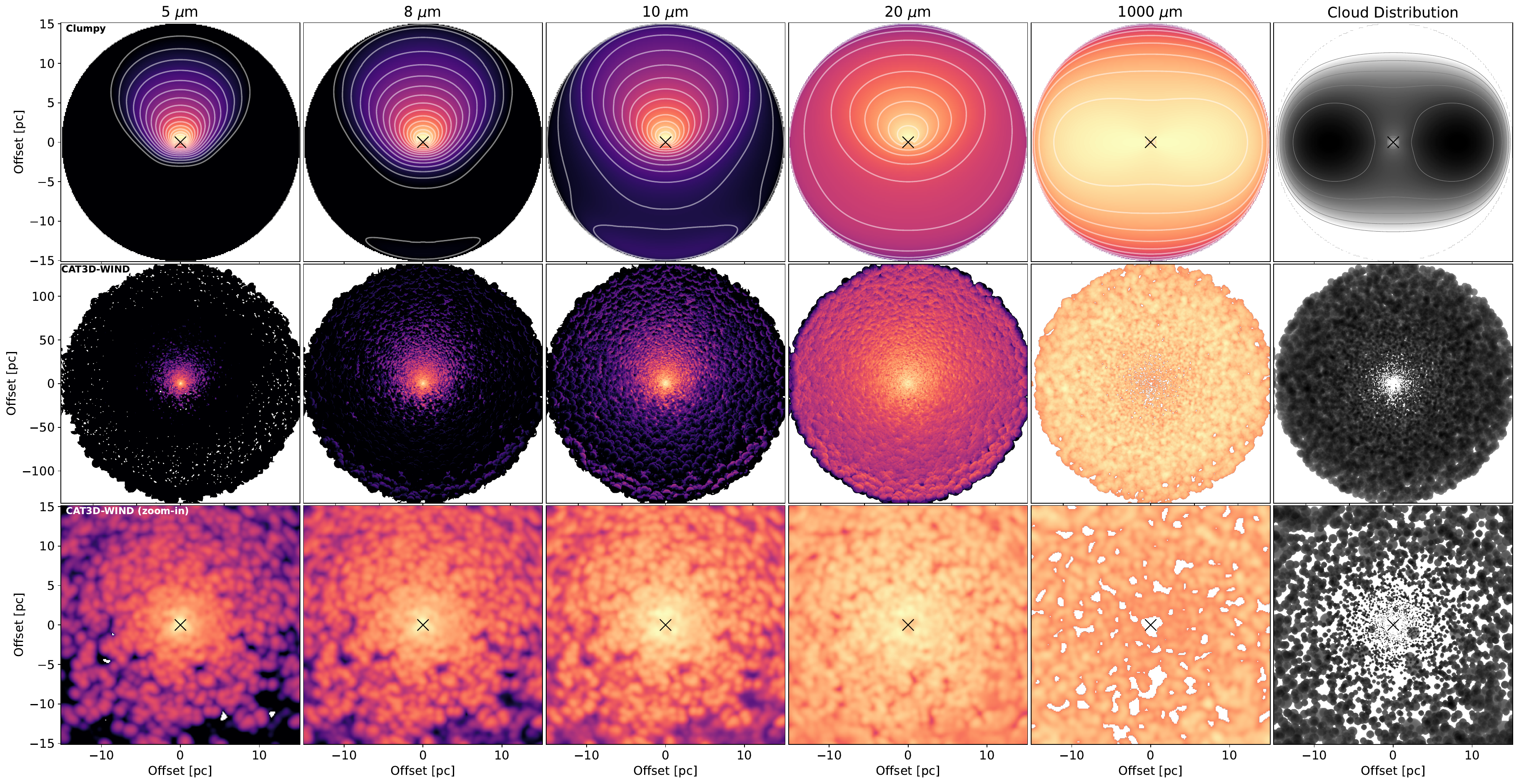}
   \caption{Model images of the clumpy (top row) and disc+wind (bottom rows) models fitted to the nuclear continuum spectrum of J1010 as they would look projected on the sky. The models correspond to the ones described by the parameters of the 50 percentile given in Tables \ref{tab:nenkova} and \ref{tab:honig}. In the case of the clumpy model, $\tau_V$=160 and Y=20 were used instead of the values shown in Table \ref{tab:nenkova} because the latter are outside the parameter range from which images are computed in HyperCat, but this have negligible impact of torus emission and cloud distribution. The last row is a zoom-in of the disc+wind model showing the same physical region as the clumpy model shown in the top row (30$\times$30 pc$^2$), since the middle row covers a FOV of 300$\times$300 pc$^2$. The physical sizes were computed by assuming a bolometric luminosity of 10$^{\rm 45}$ erg~s$^{-1}$, and in the case of the disc+wind model, a pixel scale = 0.19$\times$r$_{\rm sub}$, with r$_{\rm sub}$=0.288 pc for the largest graphite grains. From left to right, the panels correspond to torus model emission at 5, 8, 10, 20, and 1000 \micron, and to the cloud distribution considering all the mass. All the panels are shown in logarithmic color scale and they are normalized to the peak of emission. {\color {black} The black cross in the top and bottom panels indicates the position of the AGN.}}
              \label{fig:images}%
    \end{figure*}

\subsection{Model images}

\begin{table*}
\centering
\caption{Silicate feature strengths and peak wavelengths measured from the PAH-subtracted nuclear spectrum of J1010 (smoothed and interpolated) and from the torus median models fitted here.}
\begin{tabular}{lcccccc}
\hline
\hline
Silicate  & \multicolumn{2}{c}{PAH-subtracted spectrum} & \multicolumn{2}{c}{Clumpy model}  & \multicolumn{2}{c}{Disc+wind model} \\
feature   & Strength & Peak wavelength & Strength & Peak wavelength & Strength & Peak wavelength \\
\hline
9.7 \micron & 0.39 & 10.5  & 0.39 & 10.1 & 0.24 & 10.1 \\
18 \micron  & 0.15 & 18.1  & 0.10 & 17.9 & 0.13 & 18.1 \\
23 \micron  & 0.07 & 23.1  & \dots & \dots & 0.03 & 24.1 \\
\hline
\end{tabular}
\label{tab:silicates} 
\end{table*}

Figure \ref{fig:images} shows the images of the two median torus models that we fitted to the nuclear spectrum as they would look projected on the sky. The different panels correspond to the torus emission at 5, 8, 10, 20, and 1000 \micron, and to the cloud distribution considering all the mass. The innermost, directly-illuminated clouds are the hottest, therefore dominating the emission at the shortest wavelengths shown there, and as we move to longer wavelengths, we see emission from clouds at larger radii. In the top panels we show the clumpy model images generated with {\it HyperCat} \citep{Nikutta21a,Nikutta21b} for the median model fitted to the spectrum of J1010 (see Table \ref{tab:nenkova}). The torus outer radius, computed using the bolometric luminosity of the QSO2\footnote{Using Equation 1 in \citet{Nenkova08b} and R$_o$=Y$\times$R$_d$, with Y=20 instead of Y=26 (see caption of Figure \ref{fig:images}).} (log L$_{\rm bol}$=45.55; see Table \ref{tab1}), is 15 pc, and the intermediate orientation and width of the torus can be appreciated in the model images from 5 to 20 \micron. These values of the torus width and inclination are responsible for the anisotropic emission seen between 5 and 10 \micron, as we can only see the contribution from the hottest clouds between PA$\sim$45$^\circ$~and -45$^\circ$ (see \citealt{LopezRodriguez18} for an example of edge-on view of the clumpy torus model instead). The flat cloud distribution (i.e., $q$ parameter close to zero) can be seen from the corresponding cloud distribution image (top right panel) and also from the 1000 \micron~image.

In the case of the median disc+wind model, whose parameters are shown in Table \ref{tab:honig}, we generated model images of 5000$\times$5000 pixels (pixel scale=0.19$\times$r$_{\rm sub}$) that we converted to parsecs using the bolometric luminosity indicated above. The outer radius is larger than that of the clumpy torus model, of $\sim$150 pc (see middle pannels of Figure \ref{fig:images}), and because of that in the last row of Figure \ref{fig:images} we show a zoom-in of the disc+wind model in the same region covered by the clumpy model (the central 30$\times$30 pc$^2$). The almost face-on orientation of the disc+wind model is clear from the model images. This orientation produces a more isotropic emission at all wavelengths in the zoom-in part, but the outer emission is still one-sided due to obscuration from the wind (see Fig. 1 in \citealt{Hoenig17} for an edge-on view of the disc+wind model for comparison). The radial distribution of the clouds in the disc is relatively steep (i.e., more clouds toward the center; $a$=-1.7), whilst that of the wind is rather flat ($a_w$=-0.7). The flat cloud distribution of the wind can be seen from the panels showing the longer wavelengths here considered, as well as from the cloud distribution image itself. 

From the two sets of model images it is clear that the torus orientation needed to produce the spectrum shown in Fig. \ref{fig2} is far from edge-on, so we can see clumps that are close to the AGN and directly illuminated by it, while still having an obscured view of the BLR in the optical. {\color {black} It is worth emphasizing that in the context of clumpy torus models, the presence of a single cloud along the LOS, which may occur from almost any viewing angle, results in a type-2 classification. Obscuration is defined by the average number of clouds along the line-of-sight, N$_{\rm LOS}$, and their optical depth, $\tau_V$. In the case of the disc+wind median model, A$_V^{\rm LOS}$=7 mag (see Table \ref{tab:honig}), and in the clumpy torus model, A$_V^{\rm LOS}$=633 mag (see Table \ref{tab:nenkova}). Even the LOS obscuration provided by the disc+wind median model is enough to hide the BLR, since optically obscured AGN usually have A$_V\gtrsim$10 mag, and type 1.8 and 1.9 AGN A$_V\sim$3-10 mag \citep{Maiolino01,Burtscher16}. 
Indeed, from the broad H$\alpha$/H$\beta$ ratio measured from the SDSS spectrum of J1010 we estimated A$_V$>5 mag (see Section \ref{observations}), and from $\rm N_H^{CO}$ measured from CO ALMA data in the beam size of 0.25\arcsec~\citep{Audibert25}, A$_V\sim$[7,11] mag (see also Section \ref{observations}). In Tables \ref{tab:nenkova} and \ref{tab:honig} we also report the probabilities of having an obscured view of the AGN (1-P$_{\rm esc}$), which are 12\% and 89\% in the case of the median disc+wind and clumpy torus models, respectively.}

\subsection{Model spectra}

To inspect how well the fitted models reproduce the mid-infrared continuum spectrum of J1010 (solid blue line in Fig. \ref{fig2}), we can translate the fit results (i.e., models between the 16 and 84 posterior percentiles) into corresponding spectra, which are shown in orange in Fig. \ref{fig2} for the two sets of models. The black lines in the same figure are the median model and corresponding residuals. Overall, the nuclear spectrum is well reproduced by the two models, as can be seen from the residuals shown in the bottom panels of Fig. \ref{fig2}. 
The clumpy torus model reproduces better the shape of the 9.7 \micron~silicate feature, and the disc+wind model the continuum shape and the 18 \micron~silicate feature, producing residuals <15\% across the whole spectral range covered by MIRI (see bottom right panel of Fig. \ref{fig2}). Taking advantage of the capabilities of nested sampling to estimate the marginal likelihood ($Z$), we obtain $\log Z_{\rm clumpy}=-24.9$ and $\log Z_{\rm disc+wind}=-21.5$. This
points to a significant preference of the disc+wind model over the clumpy torus model for explaining the observations, under the assumption that, a priori, both models are equally probable.

{\color {black} Since the fit with the clumpy torus model shows larger residuals towards near-infrared wavelengths, we added the J, H, and Ks fluxes from the 2MASS Point Source Catalog (PSC) to Figure \ref{fig2} for comparison. We included them as upper limit because the 2MASS angular resolution is $\sim$2.5-3\arcsec~and therefore they might contain host galaxy contamination. From the residuals shown in Figure \ref{fig2} we can see that the two model SEDs show a deficit of hot dust compared with the observations, unless the yet to be determined nuclear (i.e., torus-dominated) near-infrared fluxes are well below the 2MASS upper limits. }


In order to do a quantitative comparison between the silicate features detected in the spectrum and fitted models, we measured their strength and peak wavelengths. 
The strength is defined as S$_{Si}$ = ln(F$_{\rm Si}$)$-$ln(F$_{\rm cont}$), with F$_{\rm Si}$ being the peak flux and F$_{\rm cont}$ the corresponding continuum flux at the peak wavelength, calculated by fitting a line between two anchor points bluewards and redwards of the silicate feature. The results are shown in Table \ref{tab:silicates}. Differences with the values measured from the PAH-subtracted spectrum studied in \citet{Ramos25} are due to the different method used (spline fit of the whole spectral range in \citealt{Ramos25} versus linear fit of the individual silicate features here) and mainly to the fact that here we are using an smoothed and interpolated version of the nuclear spectrum. The strength of the 9.7 \micron~silicate feature is better reproduced by the clumpy model, and in both models it peaks at bluer wavelengths (10.1 \micron) than the spectrum (10.5 \micron). However, we note that part of this discrepancy between data and models might be due to PAH subtraction, since different fits of the 11.3 and 12.5 \micron~PAHs (see inset in Fig. \ref{fig1}) resulted in silicate feature peaks that vary by $\pm$0.2 \micron. We checked that different fits of this PAH 11.3-12.5 \micron~complex with the tool developed by \citet{Donnan24} do not change the model results, since the variation in the median parameters is consistent with that associated with running the Bayesian inference several times. 

Both the peak wavelength and the strength of the 18 \micron~silicate feature are better reproduced by the disc+wind model (see Table \ref{tab:silicates}). The 23 \micron~crystaline silicate feature is not reproduced by the clumpy model, which only has three points at $\lambda$>20 \micron. The disc+wind model, which is better sampled bluewards of 30 \micron, manages to reproduce some silicate emission, but weaker and peaking at redder wavelengths than observed (see Table \ref{tab:silicates}). The disc+wind and clumpy models have different geometries but also dust composition (see captions of Tables \ref{tab:nenkova} and \ref{tab:honig} for details), which {\color {black} might} explain some of the differences between the two fits. While the clumpy torus has the typical ISM composition including silicates and graphites with maximum sizes of 0.25 \micron, the disc+wind models also include larger graphites and silicates, of up to 1 \micron, and the wind component only includes large graphites (see Fig. 1 in \citealt{GarciaBernete22}). {\color {black} As explained in \citet{Hoenig17}, this is because the wind is launched near the dust sublimation area, and therefore, their chemical compositions should be similar. This implies that the wind dust clouds in the model are devoid of silicates and small grains, avoiding the presence of strong silicate emission features. However, in reality, silicates might eventually reform in the wind (e.g., \citealt{Sarangi19}), which could explain why the 9.7 \micron~silicate emission strength in the fitted disc+wind model is weaker than the observed one.}

\section{Discussion}
\label{discussion}

In this work we have attempted to reproduce the JWST/MIRI spectrum of the QSO2 J1010, which shows the 9.7, 18, and 23 \micron~silicate features in emission, with two sets of torus models. We find that both torus models successfully reproduce the nuclear spectrum with a given combination of parameters, {\color {black} and} the disc+wind model {\color {black} is} preferred by the observations. Now we attempt to put our results in context with those found for other type-2 AGN, including quasars and Seyfert galaxies.

\subsection{Isolating nuclear dust}
\label{isolating}

      \begin{figure*}
   \centering
  \includegraphics[width=1.85\columnwidth]{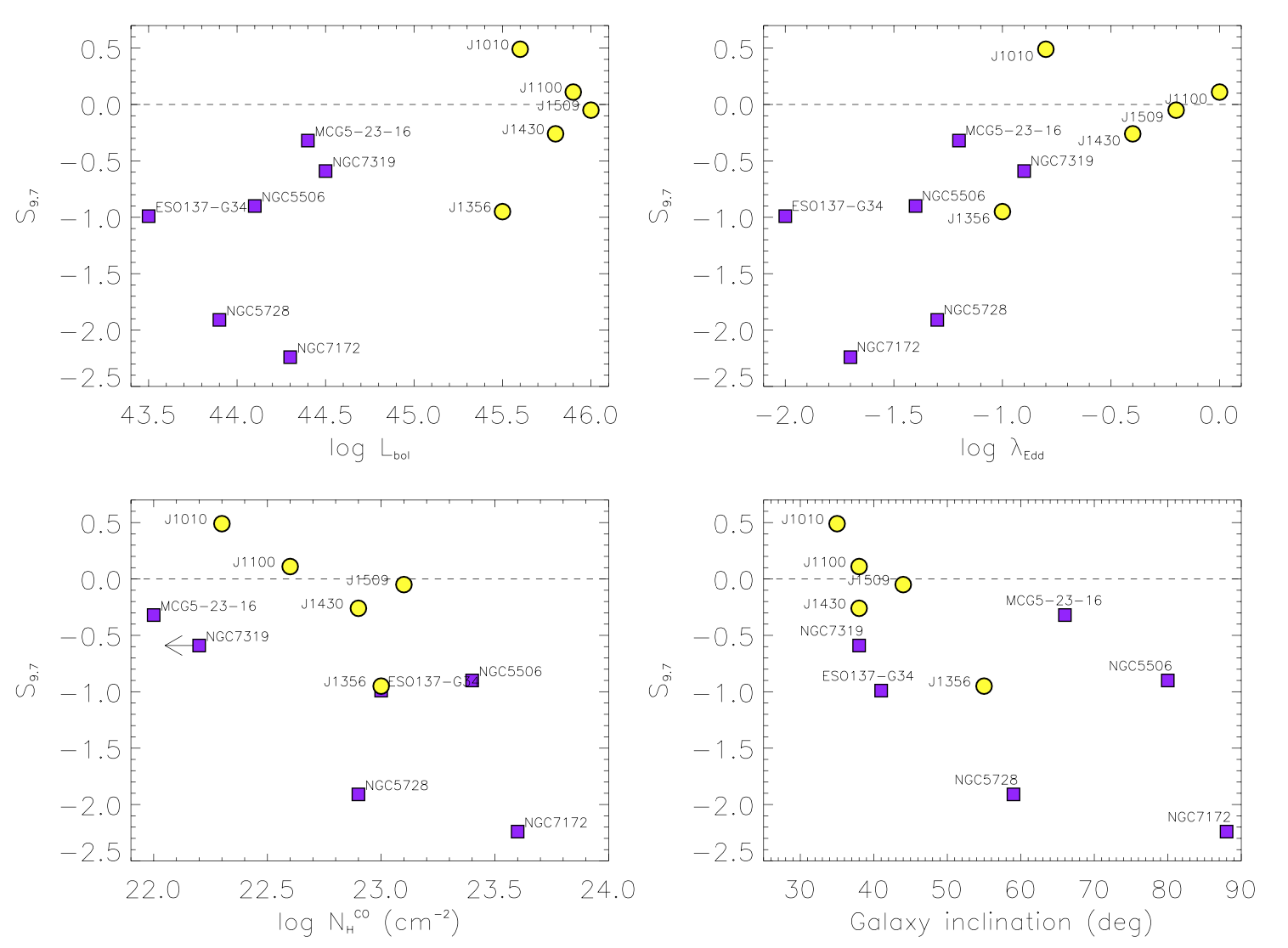}
  \caption{Strength of the 9.7 \micron~silicate feature versus bolometric luminosity, Eddington ratio, gas column density, and galaxy inclination. Yellow circles are the five QSOFEED QSO2s studied in \citet{Ramos25} and purple squares the six GATOS Seyfert 2 galaxies from \citet{GarciaBernete24}. The horizontal dashed line in all panels indicates the separation between silicate features in absorption (negative values) and in emission (positive values).}
              \label{fig:silicates}%
    \end{figure*}

Since the torus is compact in the mid-infrared (a few parsecs; \citealt{Ramos17review,Hoenig19,Gamez22,Isbell22,Haidar24}), it is not possible to distinguish between foreground (host galaxy) and intrinsic (torus) absorption in AGN from mid-infrared data alone \citep{Hatziminaoglou15}. Indeed, in dusty starbursts, mergers, and edge-on galaxies with prominent dust lanes, silicate features are often deep \citep{Gonzalez13,Hatziminaoglou15}. Using the spectral decomposition tool presented in \citet{Hernan15}, \citet{Hatziminaoglou15} reported that once they removed the host galaxy contribution from Spitzer/IRS spectra of a sample of 784 AGN, the number of type-1 (type-2) AGN with silicate features emission increased from 60\% to 80\% (12\% to 25\%). This confirms that a significant fraction of absorption features comes from cold dust in the host galaxy that does not emit in the infrared. 

The higher angular resolution of JWST/MIRI, of $\sim$0.3-0.8\arcsec, now makes it possible to get rid of a significant fraction of the host galaxy and characterize nuclear dust more accurately \citep{GarciaBernete24,Haidar24,Lopez25,Gonzalez25}. The only QSO2 in \citet{Ramos25} with a Spitzer/IRS spectrum is J1509, for which \citet{Zakamska16} reported S$_{9.7}$=-0.26 without removing the PAH bands. In \citet{Ramos25} we measured S$_{9.7}$=-0.19 (-0.05) from the JWST/MIRI nuclear spectrum without (with) PAH removal. This confirms that some of the obscuration causing the silicate absorption feature in the Spitzer/IRS spectrum originates from the host galaxy. However, in the case of this QSO2 with strong PAH features, most of the absorption is an artifact of them not being removed. The nuclear JWST spectra of J1010 and J1100 permit us to see directly illuminated dusty clouds within the torus that produce the silicate features in emission without using spectral decomposition techniques, and indeed for J1010 here we showed that its spectrum can be successfully reproduced with torus models. Finally, J1430 and J1356, despite them being a post-merger and an ongoing merger system respectively, have S$_{9.7}$=-0.26 and -0.95, which are far from the extreme values reported for edge-on and merging Seyfert galaxies even when using subarcsecond resolution data \citep{Gonzalez13,Gonzalez25,GarciaBernete24}.

Interestingly, \citet{Hatziminaoglou15} also found that 35\% of both type-1 and 2 AGN showing the silicate feature in absorption in the IRS spectra showed even deeper absorptions once the host galaxy emission was removed. This happens when the silicate feature associated with the torus is deeper than that of the host galaxy material, because the torus covering factor is high. Indeed, \citet{Gonzalez25} 
recently reported silicate features in absorption in the JWST/MIRI nuclear spectra of nearby low-to-intermediate luminosity AGN that are stronger than in the spectra of the extended emission. In their sample, the nuclear spectra of 40\% of sources cannot be reproduced by any of the currently available torus models because they are heavily obscured nuclei showing deep silicate features, ices, and aliphatic grain absorptions. New torus models including different chemistry \citep{GarciaBernete24,Reyes24,Reyes25} and larger radii \citep{GarciaBurillo21} are needed to explain the diverse AGN dust continuum emission observed by JWST \citep{Gonzalez25}.


\subsection{AGN feedback shaping nuclear dust}
\label{feedback}

The PAH-subtracted 9.7 \micron~silicate feature strengths of the five QSO2s studied in \citet{Ramos25} are shown in Figure \ref{fig:silicates} and Table \ref{tab:comparison}. We also included the {\color {black} PAH-subtracted} values reported by \citet{GarciaBernete24} for six Seyfert 2 galaxies from the GATOS sample, observed with JWST/MIRI and probing nuclear scales of $\sim$50-200 of parsecs. We chose this dataset for the comparison with the QSO2s because they were all observed with MIRI, the spectra were reduced and extracted in the same manner, and S$_{9.7}$ was determined following the same methodology (i.e., spline fit {\color {black} of the PAH-subtracted MIRI spectra to} determine the continuum). The six Seyfert 2 galaxies show silicate features in absorption (S$_{9.7}$<0) despite the small nuclear scales probed by JWST nuclear spectra. 

\begin{table}[ht]
    \caption{Properties of QSO2s and Seyfert 2 galaxies observed with JWST/MIRI and included in Figure \ref{fig:silicates}.}
    \centering
    \footnotesize
    \begin{tabular}{lccccc}
        \hline\hline
        ID & S$_{\mathrm{9.7}}$ & log L$_{\mathrm{bol}}$ & log $\lambda_{\rm Edd}$ & log N$_{\rm H}^{\rm CO}$ & i  \\
           & & (erg~s$^{-1}$) & & (cm$^{-2}$) & ($^\circ$) \\
        \hline
        QSO2s \\
        \hline
        J1010+0612  & $0.49$  & 45.6 & $-0.80$ & 22.3 & 35 \\
        J1100+0846  & $0.11$  & 45.9 & $0.00$  & 22.6 & 38 \\
        J1509+0434  & $-0.05$ & 46.0 & $-0.20$ & 23.1 & 44 \\
        J1430+1339  & $-0.26$ & 45.8 & $-0.40$ & 22.9 & 38 \\
        J1356+1026  & $-0.95$ & 45.5 & $-1.00$ & 23.0 & 55 \\
       \hline
        Seyfert 2s\\
        \hline
        MCG\,5-23-16 & $-0.32$ & 44.4 & $-1.20$ & 22.0 & 66 \\
        NGC\,7319    & $-0.59$ & 44.5 & $-0.90$ & <22.2 & 38 \\
        NGC\,5506    & $-0.90$ & 44.1 & $-1.40$ & 23.4 & 80 \\
        ESO\,137-G034& $-0.99$ & 43.5 & $-2.00$ & 23.0 & 41 \\
        NGC\,5728    & $-1.91$ & 43.9 & $-1.30$ & 22.9 & 59 \\
        NGC\,7172    & $-2.24$ & 44.3 & $-1.70$ & 23.6 & 88 \\
        \hline
    \end{tabular}
    \tablefoot{All the QSO2s' properties are from \citet{Ramos25}, except for N$_H^{\rm CO}$, which we recalculated using the CO(2-1) and CO(3-2) observations with beam sizes of $\sim$0.2-0.3\arcsec~from \citet{Audibert25}, in apertures equal to the corresponding beam sizes, and assuming $\alpha_{CO}$=4.35 M$_{\sun}$(K~km~s$^{-1}$~pc$^2$)$^{-1}$ and R$_{21}$=R$_{31}$=1. In the case of the Seyfert 2 galaxies, S$_{9.7}$ is from \citet{GarciaBernete24}, L$_{\rm bol}$ from \citet{Koss17}, and estimated from the 14-150 keV intrinsic luminosity, and $\lambda_{\rm Edd}$ and galaxy inclination from \citet{GarciaBurillo24} and \citet{Zhang24a}. For NGC\,7319, $\lambda_{\rm Edd}$ was calculated using L$_{\rm bol}$ and a black hole mass of 2.4$\times10^6$ M$_{\sun}$. N$_H^{\rm CO}$ were calculated from the ALMA CO data used in \citet{GarciaBurillo24}, which have beam sizes of $\sim$0.1-0.2\arcsec, and in \citet{Esparza24} for MCG-5-23-16 (beam size of 0.65\arcsec$\times$0.53\arcsec), using the same assumptions as for the QSO2s. For NGC\,7319, a 3$\sigma$ upper limit for N$_H^{\rm CO}$ was obtained from NOEMA CO(1-0) observations with a beam size of 0.6\arcsec$\times$0.3\arcsec~(PI: M. Pereira-Santaella), assuming $\Delta$v$_{\rm FWHM}$=250 km~s$^{-1}$ and $\Delta$v=50 km~s$^{-1}$ and using Equation 1 in \citet{Koay16}. The N$_{\rm H}^{\rm CO}$ values of the QSO2s are comparatively more beam-diluted than those of Seyfert galaxies because of the different beam sizes ($\sim$500 pc for the QSO2s versus $\sim$50-200 pc in the Seyferts).}
    \label{tab:comparison}
\end{table}

In Figure \ref{fig:silicates} we show S$_{9.7}$ versus bolometric luminosity, Eddington ratio, gas column density measured from CO, and galaxy inclination for the QSOFEED QSO2s and the GATOS Seyfert 2 galaxies. From the top panels we find moderately strong positive correlations between S$_{9.7}$ and both the bolometric luminosity and the Eddington ratio (Pearson correlation coefficients of r=0.68 and 0.70 and slopes of 0.62 and 0.93 respectively). This might be related to the relationship between Eddington ratio and torus covering factor reported by \citet{Ricci17} for $\sim$400 X-ray selected AGN in the local universe (median redshift of 0.037) with X-ray column density measurements (N$_{\rm H}^{\rm X}$). The latter authors claimed that radiation pressure on dusty gas is the main mechanism regulating the dust and gas distribution of the cicumnuclear material in AGN, as theoretically predicted \citep{Fabian06,Ishibashi18}. 
As the dust covering factor is reduced by radiation pressure, hotter material within the torus becomes exposed, making it easier to observe the silicate feature in emission {\color {black} in} luminous, higher Eddington ratio AGN. This was proposed by \citet{Maiolino07} for explaining the {\color {black} observed correlation between silicate emission strength and AGN luminosity using a sample of type-1 AGN, and} the lack of deep absorption features at the highest AGN luminosities \citep{Hatziminaoglou15,Ramos25}. 

With the aim of increasing the number of Seyfert 2 galaxies to compare with, in Table \ref{tab:s9p7} we show S$_{9.7}$ values measured from mid-infrared observations obtained with ground-based telescopes. These observations have similar angular resolution to those from JWST/MIRI, but narrower spectral range, much lower sensitivity, {\color {black} and no PAH subtraction performed. For these reasons we prefer} to keep the Seyfert 2 galaxies observed with JWST as our main comparison sample. Despite this, we can see from Table \ref{tab:s9p7} that the median S$_{9.7}$ values measured for three different samples of Seyfert 2 galaxies are -0.39, -0.28, and -0.95, i.e., larger or equal to the Seyfert 2s observed with JWST/MIRI (median value of -0.99). Indeed, only three galaxies among the four samples of Seyfert 2 galaxies considered here (over 30 targets in total, as there are a few galaxies in common among the samples) show S$_{9.7}$>0, with modest values of 0.03, 0.04, and 0.05. Therefore, silicate features in emission in Seyfert 2 galaxies are rather uncommon even when the data probe scales of tens to hundreds of parsecs, but the S$_{9.7}$ values show large scatter, going from weak or no emission to strong absorption (see Table \ref{tab:s9p7}). {\color {black} However, we note that since the ground-based spectra included in Table \ref{tab:s9p7} were not PAH-subtracted, the corresponding S$_{9.7}$ values might be biased toward deep absorption values.}

On the other hand, and despite the small number of QSO2s yet observed with JWST/MIRI, there are no QSO2 nuclear spectra showing deep silicate absorption features (S$_{9.7}$<-1). This is remarkable considering that some of them are hosted in merging and post-merger galaxies (i.e., J1356 and J1430 respectively). This {\color {black} supports} the scenario where the more luminous the AGN and the higher their Eddington ratio, the lower the torus covering factor and the more unlikely to see deep silicate absorption. In the case of the Seyfert galaxies, the covering factors are higher, but because of the torus clumpiness, it is still possible to have exposed views of directly illuminated clouds which contribute to infill the silicate feature, producing the larger scatter in S$_{9.7}$. According to this, the positive correlations seen in the top panels of Figure \ref{fig:silicates} would become more wedge-like if the number of QSO2s and Seyfert 2s observed with JWST increases, with large scatter at low L$_{\rm bol}$ and $\lambda_{\rm Edd}$ and high S$_{9.7}$ values at high L$_{\rm bol}$ and $\lambda_{\rm Edd}$.

The torus models fitted to the mid-infrared spectrum of J1010 (see Figure \ref{fig2}) have covering factors {\color {black} that are relatively 
low for an obscured AGN.} In the case of the clumpy model is C$_T$=0.45$\pm^{0.26}_{0.18}$ {\color {black} (see Table \ref{tab:nenkova})}, and in the disc+wind model C$_T$=0.66$\pm^{0.16}_{0.17}$ {\color {black} (see Table \ref{tab:honig}). We note that C$_T$ is an integrated quantity (over i and z; see Tables \ref{tab:nenkova} and \ref{tab:honig}) and this is why C$_T$(disc+wind)>C$_T$(clumpy) despite the higher $\rm A_V^{LOS}$ of the clumpy model}. These values are between those reported for type-1 (C$_T\leq$0.6) and type-2 AGN (C$_T$>0.5) using the same torus models fitted to Seyfert galaxies \citep{Ramos11b} and nearby type-1 quasars \citep{MartinezParedes17}. {\color {black} Using the same models, infrared photometry, and spectroscopic data,} \citet{Alonso11} showed that the torus covering factor decreases from C$_{\rm T}\sim$0.9-1 at low AGN luminosities (10$^{\rm 43-44}$ erg~s$^{-1}$) to C$_{\rm T}\sim$0.1-0.3 at high AGN luminosities ($\geq$10$^{\rm 45}$ erg~s$^{-1}$). The intermediate values of the covering factor that we measure for J1010, combined with the torus orientation, enable a direct view of a significant fraction of directly illuminated clumps within the torus (see Figure \ref{fig:images}). 
Torus model fits of the nuclear mid-infrared spectra of the remaining QSO2s and of the Seyfert 2 galaxies used here for comparison will be the subject of forthcoming QSOFEED and GATOS works.

In Figure \ref{fig:silicates} we also investigated the relation between S$_{9.7}$ and the gas column density measured from ALMA CO data of the QSO2s and the Seyfert galaxies (N$_{\rm H}^{\rm CO}$). We find a moderately strong negative correlation in this case (r=-0.60 and slope of -1.00), with the lowest column density targets generally showing S$_{9.7}$>-0.5 and the highest column density ones showing S$_{9.7}\lesssim$-0.5. The targets with log N$_{\rm H}^{\rm CO}\sim$ 23 cm$^{-2}$ cover a wide range of S$_{9.7}$ values, in some cases possibly having a fraction of the silicate absorption coming from the host galaxy because they are edge-on galaxies (NGC\,5506 and NGC\,7172; see bottom right panel of Figure \ref{fig:silicates}) or ongoing merger systems as J1356 \citep{Ramos22}. Indeed, we also find a moderately strong correlation between S$_{9.7}$ and galaxy inclination (using cos(i); r=0.69 and slope of 2.10). However, we checked that if we exclude the two edge-on galaxies, the correlations shown in the top panels of Figure \ref{fig:silicates} hold (r=0.70 and 0.62), whereas those in the bottom panels become weaker (r=-0.38 and 0.51). The negative trend between S$_{9.7}$ and gas column density was first reported by \citet{GarciaBernete24} for the GATOS Seyfert 2s and a small sample of ULIRGs observed with JWST/MIRI and Spitzer/IRS respectively, but using N$_{\rm H}^{\rm X}$ instead of N$_{\rm H}^{\rm CO}$ (r=0.68 and slope of 0.29).

  \begin{figure}
   \centering
  \includegraphics[width=1.0\columnwidth]{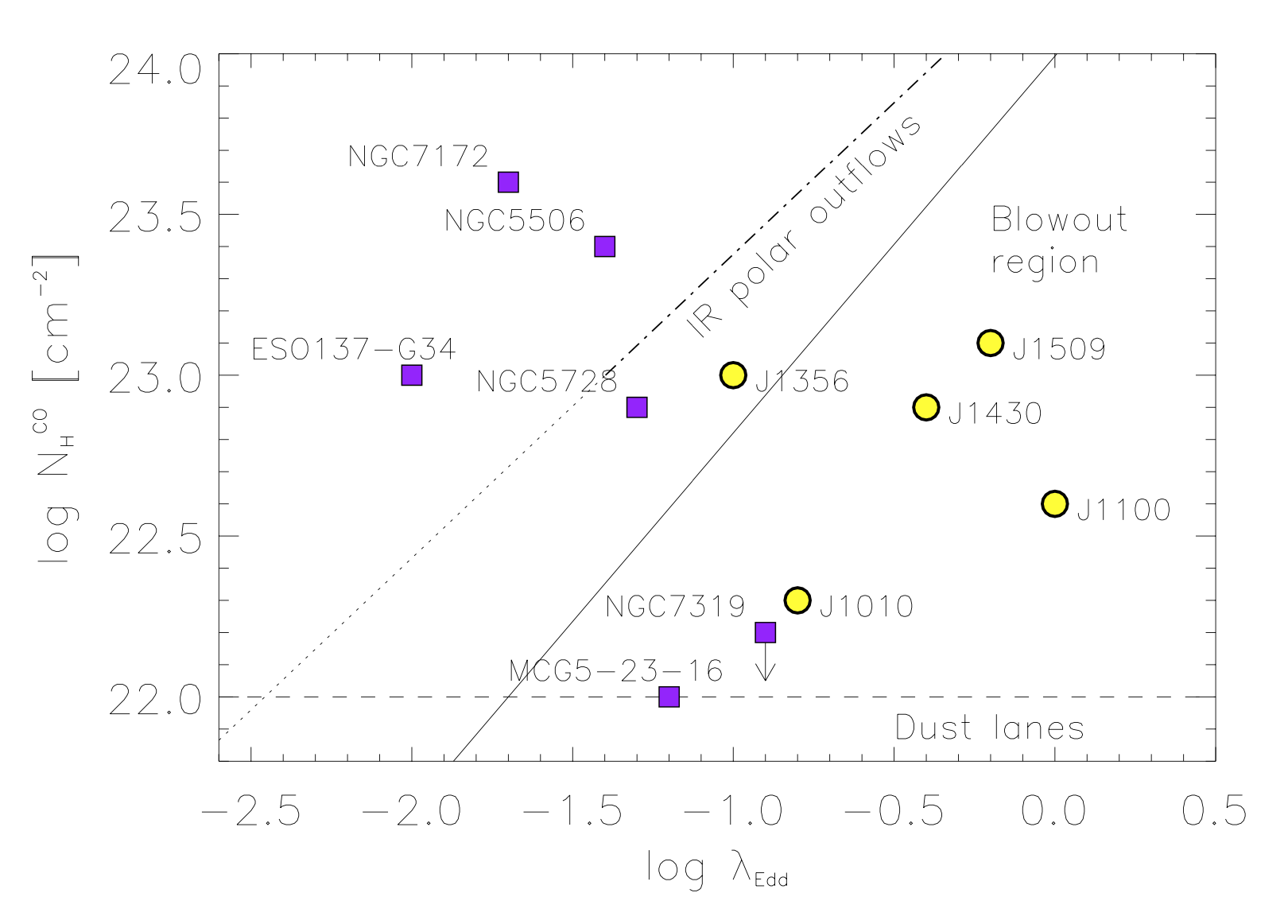}
  \caption{Column density measured from CO versus Eddington ratio. The blowout region is where radiation pressure pushes away the
obscuring material ($\lambda_{\rm Edd}$>$\lambda_{\rm Edd}^{\rm eff}$(N$_H$)), from \citet{Ricci17}. The dot-dashed line from \citet{Venanzi20} indicates the limit where the AGN radiation acceleration balances gravity and IR radiation pressure dominates, giving rise to polar dusty outflows. The dotted line is an extrapolation of this limit to lower column densities, and the horizontal dashed line is the approximate upper limit for absorption due to dust lanes \citep{Ricci17}. Yellow circles are the five QSO2s from \citet{Ramos25} and purple squares are the Seyfert galaxies from \citet{GarciaBernete24}.}
              \label{fig:nh_edd}%
    \end{figure}

Finally, we checked where the QSO2s lie in the Eddington ratio-column density diagram (see Figure \ref{fig:nh_edd}; \citealt{Fabian06,Ricci17,GarciaBernete22}), using N$_{\rm H}^{\rm CO}$ values as in \citet{AlonsoHerrero21} and \citet{GarciaBurillo24}. The blowout or ``forbidden'' region, rightward of the solid line, is the area of the diagram where radiation pressure pushes away the obscuring material ($\lambda_{\rm Edd}$>$\lambda_{\rm Edd}^{\rm eff}$(N$_H$); \citealt{Ricci17}), resulting in a small amount of {\color {black} low-luminosity} AGN there \citep{Ricci17,GarciaBurillo24}. The dot-dashed line shown in Figure \ref{fig:nh_edd}, extrapolated to lower column densities (dotted line), is the theoretical limit where AGN radiation balances gravity and infrared radiation pressure on dust dominates, from \citet{Venanzi20}. According to these predictions, no targets should be present in the blowout region at log N$_H$>22 cm$^{-2}$, and polar dusty winds should be present when the objects are close to the dot-dashed/dotted line. Remarkably, four of the five QSO2s are in the blowout region of the diagram, which is at odds with the small percentage of X-ray selected AGN reported by \citet{Ricci17} for this region, of 1.4$\pm^{0.7}_{0.5}$\%, and by \citet{GarciaBurillo24} using CO-detected nearby AGN with high angular resolution ALMA observations (2\%). 
This suggests that, whereas Seyfert galaxies retain long-lived obscuring clouds, our optically selected QSO2s are actively clearing gas and dust from their nuclear regions, leading to reduced covering factors. Indeed, here we showed that the mid-infrared spectrum of J1010 is best reproduced with disc+wind torus models having a covering factor of {\color {black} C$_T$=0.66$\pm^{0.16}_{0.17}$}. From previous work on the five QSOFEED QSO2s we know that they are experiencing strong AGN feedback in different gas phases. For example they show molecular gas outflows detected using ALMA CO(2-1) observations \citep{Ramos22,Audibert23,Audibert25}, with radii of 0.4-2.4 kpc and mass rates of 5-60 M$_{\sun}$~yr$^{-1}$. 
\citet{Lansbury20} reported similar findings for a sample of infrared-selected heavily dust-reddened quasars (red quasars), which were also in the blowout region of the N$_H^X-\lambda_{\rm Edd}$ diagram. Recently, using MATISSE interferometry data, \citet{Drewes25} found that the high Eddington sources I Zw 1 and H0557-385 do not show pc-scale polar emission but rather an equatorial extension and a possible puffed-up component at 3-5 \micron. This was interpreted as due to radiation pressure blowing out dust close to the equatorial plane and exposing the wind-launching region, 
giving rise to stronger silicate features than in low-Eddington ratio AGN. 

Finally, for low-Eddington ratio sources, high-angular resolution ALMA CO observations of a sample of 70 local AGN revealed a dichotomy in the location of the galaxies in the ``permitted'' region of the N$_H^{\rm CO}-\lambda_{\rm Edd}$ diagram \citep{GarciaBurillo24}. This dichotomy depends on whether they belong to the AGN build-up or feedback branches (high and low nuclear molecular gas concentrations respectively), indicating that radiation pressure on dusty gas also determines the position of Seyfert galaxies in different (permitted) areas of this diagram. As we can see from Figure \ref{fig:nh_edd}, the Seyfert 2 galaxies from GATOS \citep{GarciaBernete24} are also in the ``permitted'' region, with MCG-5-23-16 and NGC\,7319 having log N$_H\lesssim$22 cm$^{-2}$. These low column densities can be explained by host galaxy dust lanes \citep{Fabian06,Ricci17}, although they both show nuclear molecular gas deficits in the CO data \citep{Esparza24} that could be due to AGN feedback \citep{GarciaBurillo24}. These two galaxies show the weakest silicate absorption features in \citet{GarciaBernete24}. 

Therefore, based on our own and previous results, the ``forbidden'' {\color {black} or} blowout region of the diagram is avoided by low-luminosity AGN, which have long-lived tori with higher covering factors, but not by {\color {black} optically} obscured quasars and red quasars, which likely transition across the diagram {\color {black} on shorter timescales than Seyferts} due to the stronger AGN feedback that they experience.

\section{Conclusions}

The JWST/MIRI mid-infrared spectrum of the central 0.5-1.1 kpc of the QSO2 J1010, which is AGN-dominated and shows 9.7, 18, and 23 \micron~silicate features in emission, {\color {black} is successfully reproduced by torus models without including direct emission from the accretion disc. This is the first MIRI spectrum of a QSO2 fitted with torus models.} Based on the marginal likelihood and residuals from the fits, we find that the 
disc+wind model is preferred by the observations. This model 
has an {\color {black} intermediate covering factor ($\rm C_T=0.66\pm^{0.16}_{0.17}$) and low inclination ($\rm i=13^\circ\pm^{7^\circ}_{6^\circ}$)}, leading to exposed views of the directly illuminated clumps that produce the silicate features in emission, {\color {black} while still obscuring the BLR and accretion disc. }

The PAH-subtracted silicate features of the five QSO2s observed with JWST/MIRI, including J1010, range from absorption to emission (stengths of S$_{9.7}$=[-1.0,0.5]; \citealt{Ramos25}), with a median value of -0.05, whereas the six Seyfert 2 galaxies from \citet{GarciaBernete24}, also observed with JWST/MIRI and probing nuclear scales of $\sim$50-200 pc, show silicate features in absorption (S$_{9.7}$=[-2.2,-0.3], with a median value of -0.99). Combining the two JWST/MIRI datasets, we show that there is a moderately strong positive correlation between S$_{9.7}$ and both log $\lambda_{\rm Edd}$ and log L$_{\rm bol}$ (correlation coefficients of r=0.70 and 0.68 respectively). 

This, together with the position of the QSO2s and Seyfert 2s in the blowout region of the Eddington ratio-column density diagram, indicates that the difference between the nuclear obscuration properties of QSO2s and Seyfert 2s is likely driven by radiation pressure on dusty gas. QSO2s {\color {black} would be} experiencing a relatively short blowout phase, leading to smaller torus covering factors and hence more exposed view of nuclear gas and dust, whilst Seyfert 2 galaxies have longer-lived and higher covering factor tori, making it difficult to observe silicate features in emission.

\begin{acknowledgements}
This work is based on observations made with the NASA/ESA/CSA James Webb Space Telescope. The data were obtained from the Mikulski Archive for Space Telescopes at the Space Telescope Science Institute, which is operated by the Association of Universities for Research in Astronomy, Inc., under NASA contract NAS 5-03127 for JWST and from the European JWST archive (eJWST) operated by the ESAC Science Data Centre (ESDC) of the European Space Agency. These observations are associated with program GO 3655. This work is based on observations carried out under project number W22CN with the IRAM NOEMA Interferometer. IRAM is supported by INSU/CNRS (France), MPG (Germany) and IGN (Spain). CRA thanks Prof. Andrew Fabian for useful discussions. 
CRA and AA acknowledge support from the Agencia Estatal de Investigaci\'on of the Ministerio de Ciencia, Innovaci\'on y Universidades (MCIU/AEI) under the grant ``Tracking active galactic nuclei feedback from parsec to kiloparsec scales'', with reference PID2022$-$141105NB$-$I00 and the European Regional Development Fund (ERDF). AAR acknowledges support from the Agencia Estatal de Investigaci\'on of the Ministerio de Ciencia, Innovaci\'on y Universidades (MCIU/AEI) under the grant with reference PID2022$-$136563NB$-$I0 and the European Regional Development Fund (ERDF). IGB is supported by the Programa Atracci\'on de Talento Investigador ``C\'esar Nombela'' via grant 2023-T1/TEC-29030 funded by the Community of Madrid. ELR thanks support by the NASA Astrophysics Decadal Survey Precursor Science (ADSPS) Program (NNH22ZDA001N-ADSPS) with ID 22-ADSPS22-0009 and agreement number 80NSSC23K1585. SH acknowledges support through UK Research and Innovation (UKRI) under the UK government’s Horizon Europe Funding Guarantee (EP/Z533920/1, selected in the 2023 ERC Advanced Grant round) and an STFC Small Award (ST/Y001656/1). AA acknowledges funding from the European Union (WIDERA ExGal-Twin, GA 101158446). SGB acknowledges support from the Spanish grant PID2022-138560NB-I00, funded by MCIN/AEI/10.13039/501100011033/FEDER, EU. MPS acknowledges support under grants RYC2021-033094-I, CNS2023-145506 and PID2023-146667NB-I00 funded by MCIN/AEI/10.13039/501100011033 and the European Union NextGenerationEU/PRTR. AAH acknowledges support from grant PID2021-124665NB-I00  funded by MCIN/AEI/10.13039/501100011033 and by ERDF A way of making Europe. OGM acknowledges financial support by the UNAM PAPIIT project IN109123 and CONAHCyt Ciencia de Frontera project CF-2023-G-100. {\color {black} We finally thank the annonymous referee for their constructive  report.} 
\end{acknowledgements}

\bibliographystyle{aa} 
\bibliography{aa57323-25} 

\onecolumn
\begin{appendix} 
\section{Posterior distributions from the fits}\label{appendix}

  \begin{figure}[!h]
   \centering
  \includegraphics[width=0.9\columnwidth]{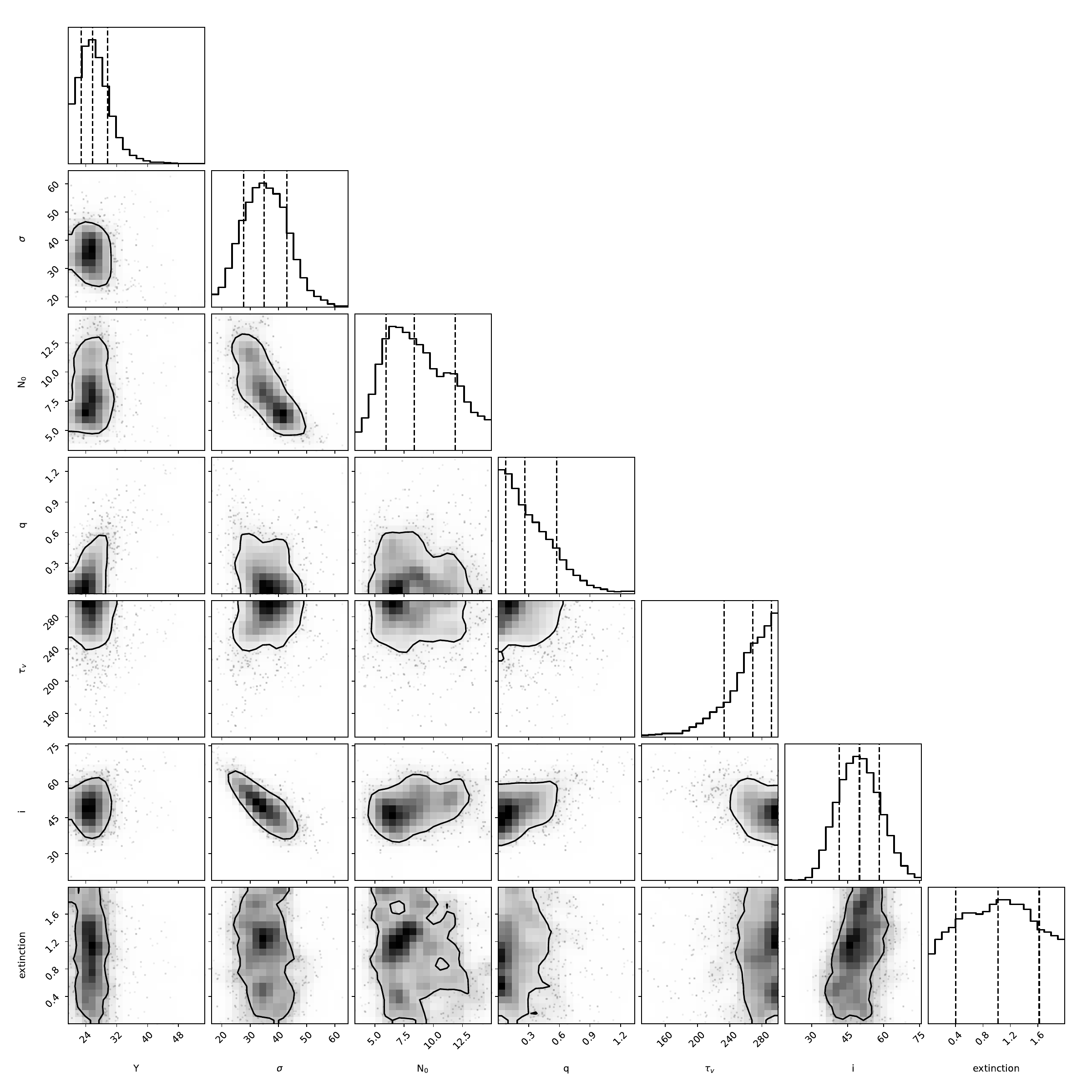}
   \caption{1D and 2D posterior distributions of the model parameters resulting from the fit of J1010 with the clumpy torus model of \citet{Nenkova08a}. The vertical dashed lines correspond to the percentiles at 16, 50, and 84, whose values are reported in Table \ref{tab:nenkova}.}
              \label{posteriors_nenkova}%
    \end{figure}

  \begin{figure}[!h]
   \centering
  \includegraphics[width=1.0\columnwidth]{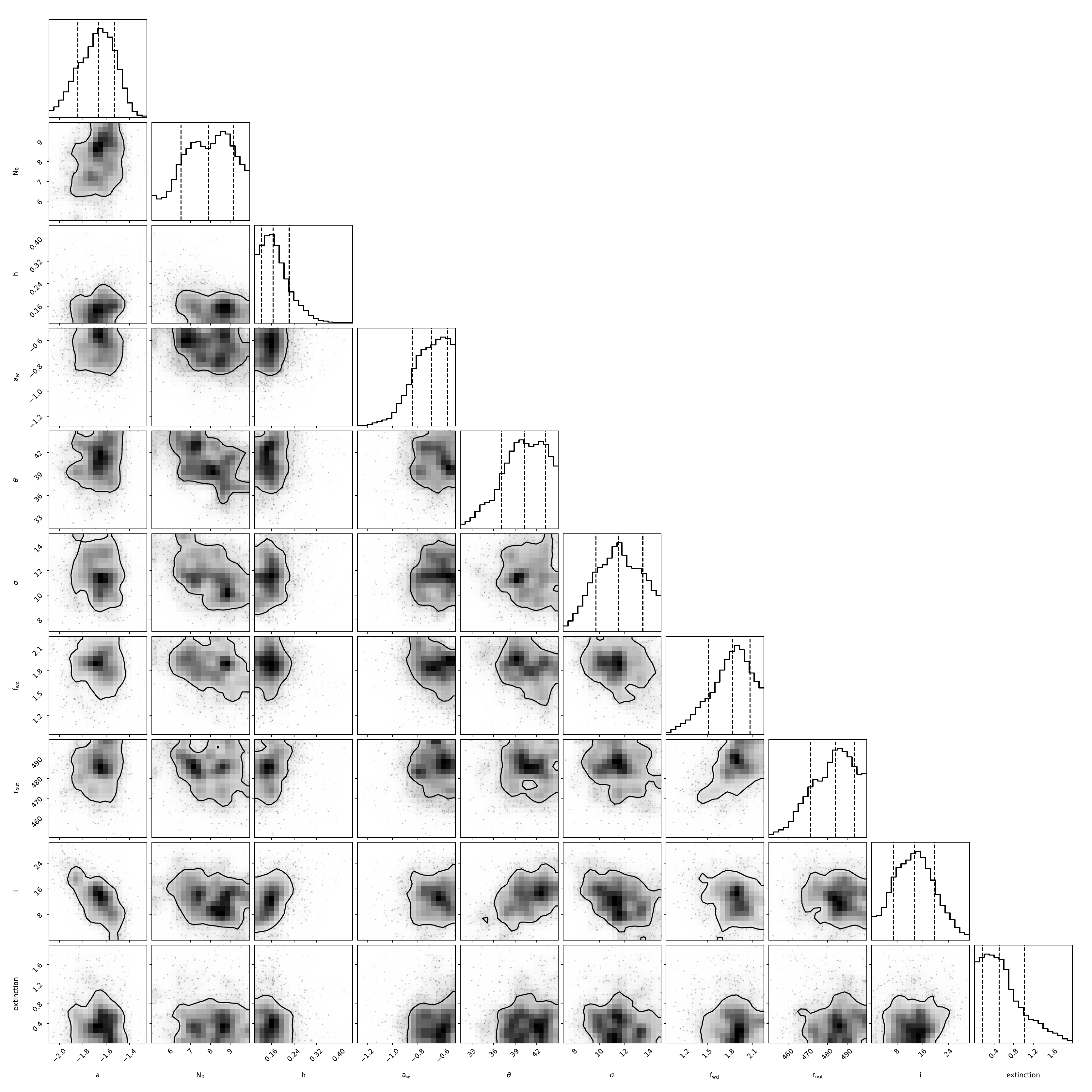}
   \caption{1D and 2D posterior distributions of the model parameters resulting from the fit of J1010 with the disc+wind model of \citet{Hoenig17}. The vertical dashed lines correspond to the percentiles at 16, 50, and 84, whose values are reported in Table \ref{tab:honig}.}
              \label{posteriors_hoenig}%
    \end{figure}

\section{Silicate {\color {black} feature} strengths of different samples of type-2 AGN}\label{appendixb}
    
    \begin{table}[!h]
\centering
\caption{Silicate {\color {black} feature} strengths of different samples of type-2 AGN measured from subarcsecond resolution mid-infrared spectra.}
\label{tab:s9p7}
\begin{tabular}{lccccc}
\hline\hline
Sources (n) & Data & \multicolumn{3}{c}{S$_{9.7}$} & Ref. \\
\cline{3-5}
 &  & range & mean & median & \\
\hline
QSO2 (5)               & JWST/MIRI       & [-0.95,0.49] & -0.13$\pm$0.53 & -0.05 & a \\
Sy2 (6)                & JWST/MIRI       & [-2.24,-0.32] & -1.16$\pm$0.76 & -0.99 & b \\
Sy2 (11)               & GTC/CanariCam   & [-1.11,0.04]  & -0.47$\pm$0.38 & -0.39 & c \\
Sy2 (10)               & VLT/VISIR       & [-1.03,0.05]  & -0.31$\pm$0.32 & -0.28 & d \\
Sy2 (16)               & Gemini/T-ReCS   & [-3.78,-0.14] & -1.25$\pm$0.97 & -0.95 & e \\
\hline
\end{tabular}
\tablefoot{Only QSO2s and type 2, 1.9, and 1.8 Seyferts were considered. The angular resolution of the CanariCam, VISIR, and T-ReCS data is similar to MIRI ($\sim$0.3-0.6\arcsec), but the spectral range covered is narrower ($\sim$8-13 \micron), and the sensitivity is much lower. Refs: (a) \citet{Ramos25}, 
(b) \citet{GarciaBernete24}, (c) \citet{Alonso16}, (d) \citet{Honig10}, (e) \citet{Gonzalez13}.}
\end{table}

\end{appendix}


\end{document}